\documentclass[useAMS]{mn2e}
\usepackage{lscape}
\usepackage{rotating}
\usepackage{amssymb}
\usepackage{float}
\usepackage{hyperref}
\usepackage{newtxtext,newtxmath}

\def\msun{\hbox{M$_\odot$}}

\def\t4{\hbox{t$_{\rm 4}$}}

\def\cm3{\hbox{cm$^{-3}$}}

\input psfig.sty
\input epsf.sty
\usepackage{graphicx}
\usepackage{epsfig}
\title[eMSTOs in Open Clusters]
{Extended Main Sequence Turnoffs in Open Clusters as Seen by Gaia: I. NGC 2818 and the Role of Stellar Rotation}
\author[Bastian et al.] {N. Bastian$^1$, S. Kamann$^{1}$,  I. Cabrera-Ziri$^{2}$\thanks{Hubble Fellow}
, C. Georgy$^{3}$, S. Ekstr\"om$^{3}$,  \newauthor C. Charbonnel$^{3,4}$, M. de Juan Ovelar$^{1}$,  and C. Usher$^{1}$ \\
$^{1}$Astrophysics Research Institute, Liverpool John Moores University, 146 Brownlow Hill, Liverpool L3 5RF, UK\\
$^{2}$ Harvard-Smithsonian Center for Astrophysics, 60 Garden Street, Cambridge, MA 02138, USA\\
$^{3}$ Department of Astronomy, University of Geneva, Chemin des Maillettes 51, 1290, Versoix, Switzerland\\
$^{4}$ IRAP, UMR 5277, CNRS and UniversitŽ de Toulouse, 14, avenue ƒdouard Belin, 31400 Toulouse, France\\
}
\date{Accepted. Received; in original form}
\pagerange{\pageref{firstpage}--\pageref{lastpage}}
\pubyear{2015}
\begin{document}
\maketitle
\label{firstpage}
\begin{abstract}
We present an analysis of the relatively low mass ($\sim2400$~\msun), $\sim800$~Myr, Galactic open cluster, NGC~2818, using Gaia DR2 results combined with VLT/FLAMES spectroscopy.  Using Gaia DR2 proper motions and parallax measurements we are able to select a clean sample of cluster members.  This cluster displays a clear extended main sequence turn-off (eMSTO), a phenomenon previously studied mainly in young and intermediate age massive clusters in the Magellanic clouds.  The main sequence of NGC~2818 is extremely narrow, with a width of $\sim0.01$ magnitudes (G$_{\rm BP} - $ G$_{\rm RP}$), suggesting very low levels of differential extinction.  Using VLT/FLAMES spectroscopy of 60 cluster members to measure the rotational velocity of the stars (Vsini) we find that stars on the red side of the eMSTO have high Vsini ($>160$~km/s) while stars on the blue side have low Vsini ($<160$~km/s), in agreement with model predictions.  The cluster also follows the previously discovered trend between the age of the cluster and the extent of the eMSTO.  We conclude that stellar rotation is the likely cause of the eMSTO phenomenon.  

\end{abstract}

\section{Introduction}
\label{sec:intro}

The extended main sequence turn-off (eMSTO) phenomenon is an observational feature which has been found in young and intermediate age (up to $\sim2$~Gyr) clusters in the Large and Small Magellanic clouds, where the MSTO of the cluster in colour-magnitude diagrams is more extended than would be expected for a simple stellar population (after accounting for binaries, extinction, and photometric uncertainties).  The origin of the eMSTO feature was initially suggested to be due to a large age spread (hundreds of Myr) within these clusters (e.g., Mackey \& Broby Nielson~2007; Milone et al. 2009; Goudfrooij et al. 2014), although subsequent work has shown that a stellar evolutionary effect, likely stellar rotation, was the most probable cause (e.g., Bastian \& de Mink~2009; Brandt \& Huang~2015b; Niederhofer et al.~2015; Cabrera-Ziri et al.~2016).   Actual age spreads, as the origin of the eMSTO, are disfavoured due to the observed relation between the extent of the inferred age spread and the age of the clusters (Niederhofer et al.~2015).\footnote{See Bastian \& Lardo~(2018) for further discussion of the evidence against actual age spreads within massive clusters.}  Spectroscopic studies of stars on the eMSTO in these clusters have found evidence that indeed stellar rotation is directly linked to the position within the MSTO (Dupree et al.~2017; Kamann et al.~2018, Marino et al.~2018).

There have been suggestions that the width of the eMSTO is linked to the cluster mass, with higher mass clusters displaying broader spreads (e.g., Conroy \& Spergel~2011; Goudfrooij et al.~2014).  However, the eMSTO feature has been found in clusters with masses down to $2-5 \times10^3$~\msun\ (Piatti \& Bastian~2016), and statistical analyses found that cluster age had a much larger correlation with eMSTO widths than mass (Niederhofer et al.~2016).  Recent work by Martocchia et al.~(2018) found that NGC~1978, a massive ($\sim3\times10^5$~\msun), $2$~Gyr cluster does not host an eMSTO, adding further support that age, and not mass, is the controlling parameter (see also Georgy et al.~2018).


Most studies of eMSTOs in clusters have focussed on massive LMC/SMC clusters with Hubble Space Telescope observations taking advantage of the lack of background/foreground confusion, high membership probabilities, and large numbers of stars on the MSTO (10s to 100s) that allow statistical analyses.  While there have been some attempts to find and study the eMSTO phenomenon in open clusters in the Galaxy (e.g., Brandt \& Huang~2015a) such studies have been limited due to the low number of stars on the MSTO, or in the case of Trumpler 20 (Platais et al.~2012) no eMSTO was found (once differential extinction was taken into account).  There have been previous suggestions in open clusters that a dual or extended main sequence turn-off could be due to rapid rotators within the clusters (Twarog~1983; Twarog et al.~2015). If clean samples of cluster members can be found in open clusters, it opens the way for detailed follow up studies, as these stars are $3-8$~magnitudes brighter than their LMC/SMC counterparts.  Additionally, by studying Galactic open clusters, the role of metallicity can be explored as, to date, only SMC/LMC metallicity environments have been studied.

Gaia DR2 (Gaia Collaboration et al. 2016, 2018) has opened the possibility of an entirely new perspective on the problem, by allowing the study of eMSTOs (and the related phenomenon of split main-sequences - e.g., Milone et al.~2016) in relatively low mass ($\lesssim3000$~\msun) open clusters in the Galaxy.  This is due to the ability to select extremely clean samples of cluster members, mitigating the effects of contamination from fore/back-ground stars.

In the present paper we exploit Gaia DR2 and archival VLT/FLAMES spectroscopy to discover and analyse the eMSTO in the $\sim800-900$~Myr open cluster, NGC~2818.  This is a first in a series of papers focussed on the eMSTO phenomenon in open clusters.

\section{Observations and Methods}
\label{sec:obs}

\subsection{Selection of cluster members}

\begin{figure*}
\centering
\includegraphics[width=14cm]{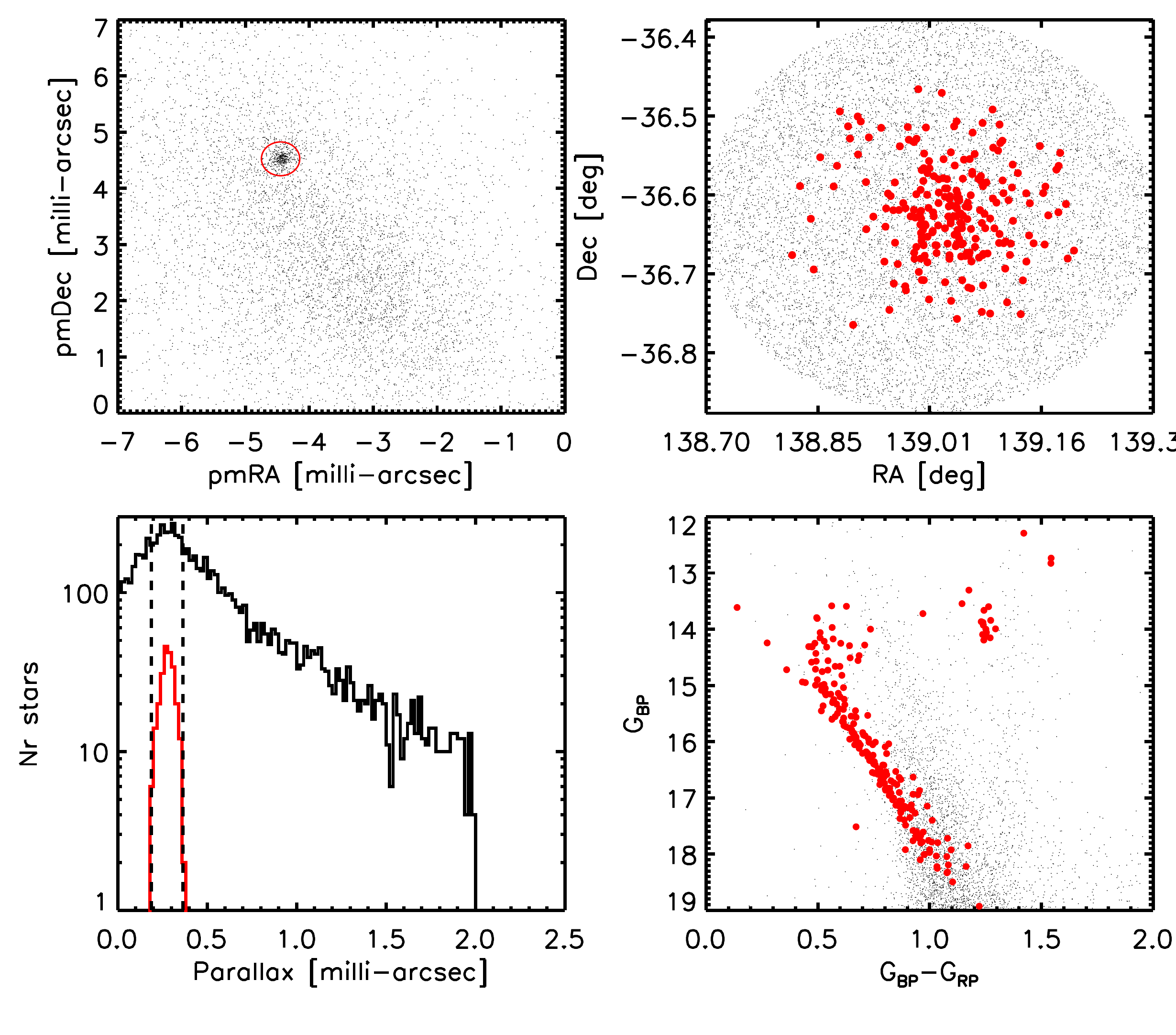}
\caption{{\bf Upper left:} Proper motion of all stars in the Gaia catalogue within 15' of the centre of NGC 2818.  A clear over-density of points belonging to cluster is seen.  The circle shows the first selection of cluster members.  {\bf Upper right:} The spatial distribution of all stars selected with cluster members highlight as filled (red) points.  {\bf Bottom left:} The parallax distribution of all stars (upper curve, black) and selected cluster members (lower curver, red). The vertical dashed lines show the cuts applied to select cluster members.  {\bf Bottom right:}  The CMD (in Gaia blue and red passbands) of all stars in the field (points) and cluster members (filled red circles).  An extended main sequence turn-off is clearly visible at G$_{\rm BP} \approx 14.5$~mag. }
\label{fig:selection}
\end{figure*}

We searched the Gaia DR database for all sources within 12' of the centre of NGC~2818 (Gaia Collaboration et al. 2016, 2018).  We then looked at the resulting proper motion map of these stars and a clear overdensity could be seen at pmRA$=-4.425$ and pmDec$=4.532$ (top left panel of Fig.~\ref{fig:selection}).  We used all stars with proper motions values of 0.3~milli-arcseconds of this centre for our first selection.  These stars showed a clear peak in the parallax distribution at 0.281~milli-arcseconds, corresponding to a distance of $3.56$~kpc which we will adopt throughout this work\footnote{This does not include the $-0.029$ milli-arsecond zeropoint offset found by Lindegren et al.~(2018) for Quasars in DR2.}.  We then made a further selection in parallax, $0.1885 \le$ parallax $\le 0.365$ (shown as the vertical dashed lines in the lower left panel of Fig.~\ref{fig:selection}) and applied a radial cut of 10' to obtain our final member sample.  Their spatial distribution and colour-magnitude diagrams (using the blue and red passband photometry from Gaia) are shown in the top-right and bottom-right panels of Fig.~\ref{fig:selection}, respectively.\footnote{A list with the photometric properties of all stars selected with the above criteria is available through the electronic journal website.}

As already evident from Fig.~\ref{fig:selection}, NGC~2818 displays a clear extended main sequence turn-off.

\subsection{Rotational velocities}

\begin{figure}
\centering
\includegraphics[width=8cm]{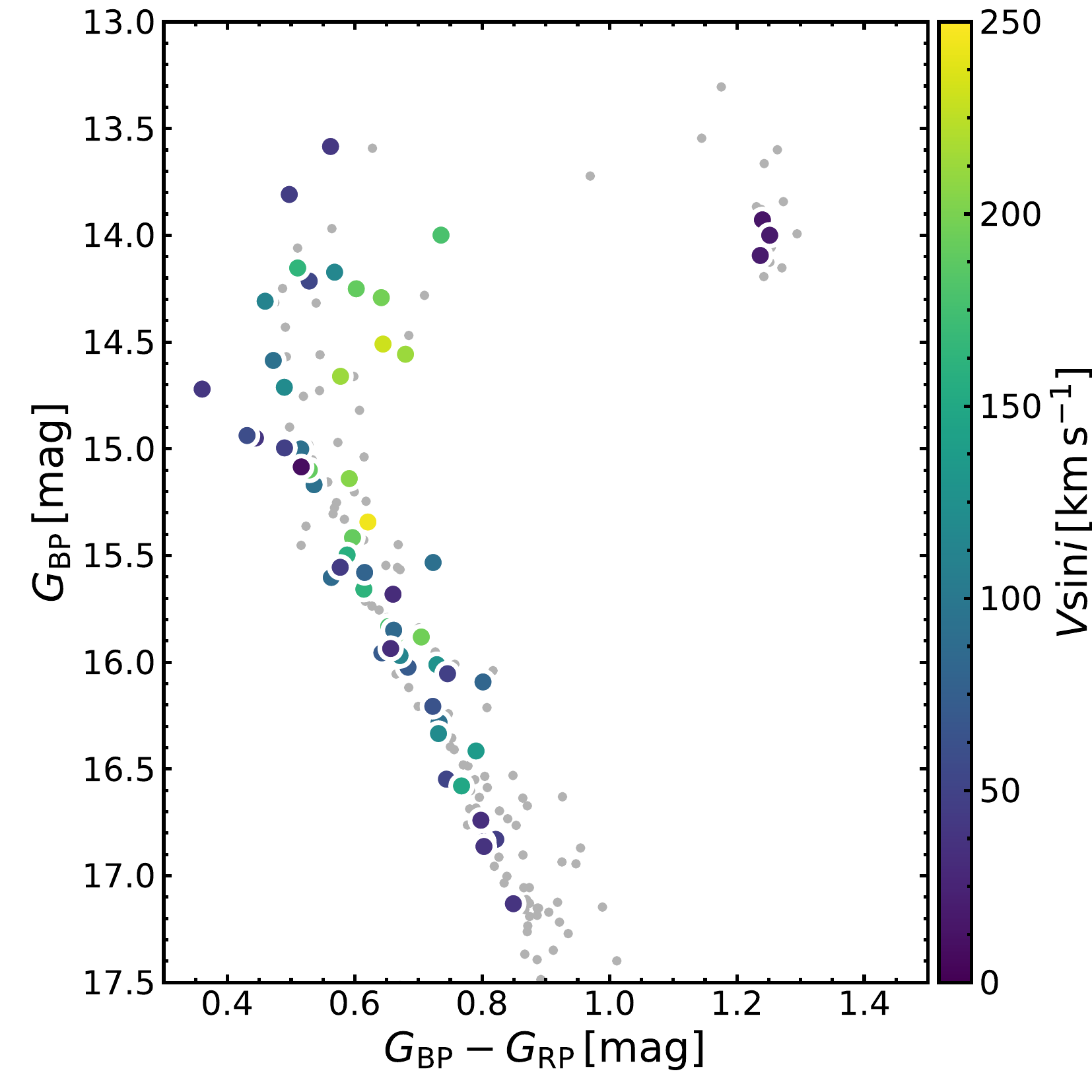}
\caption{The CMD of NGC~2818 with colour-coding showing the measured Vsini of each star.  Rapidly rotating stars preferentially lie on the red side of the eMSTO while slower rotating stars tend to be on the blue side, consistent with expectations if rotation is the cause of the eMSTO phenomenon. }
\label{fig:cmd_rotation}
\end{figure}

Spectroscopic observations of main-sequence (MS) and turn-off (TO) stars in NGC~2818 are publicly available in the ESO archive. Observations were conducted in 2014 March using the GIRAFFE multi-object spectrograph in its MEDUSA mode under the programme 093.D-0868(A) (PI: Simunovic).  The employed setups are HR05 and HR09 that offer a nominal resolution of R$\sim$20\ 250 and 18\ 000 at the central wavelengths of 4471 and 5258\AA, respectively. Stars were observed for 2$\times$45 min in each grating, leading to a signal-to-noise (S/N) of $\sim$100 and $\sim$ 20 (per single exposure) for the brightest and the faintest star in the sample, respectively.  We retrieved the extracted, wavelength calibrated, one-dimensional (1D) spectra from the Phase 3 ESO archive. The reduced MEDUSA 1D-spectra come with error estimates and signal-to-noise ratio per wavelength bin and the wavelength scale is in the heliocentric reference system.  A total of 60 stars from 093.D-0868(A) are included in the Gaia DR2 photometric catalog, and are  shown in Fig.~\ref{fig:cmd_rotation}.

To determine rotational velocities we performed a $\chi ^2$ minimization procedure between synthetic and the observed spectra.  Atmospheric parameters were derived as follows. We adopted the effective temperatures, T$_{\rm eff}$, from Gaia DR2. Surface gravities were derived from DR2 parallaxes, using the fundamental relation:

$$ \log (\frac{g}{g_{\odot}})= \log(\frac{M}{M_{\odot}}) +4 \log(\frac{T_{\rm eff}}{T_{\odot}}) - \log (\frac{L}{L_{\odot}})$$

We use V magnitudes from Stetson~(2000)\footnote{{\url{http://www.cadc-ccda.hia-iha.nrc-cnrc.gc.ca/en/community/STETSON/standards/}}}, bolometric corrections are based on Flower et al.~(1996) and Torres~(2010), and solar magnitudes from Bessell, Castelli, \& Plez~(1998).  The masses were set to 1.6~M$_{\odot}$ and the resolution is set to the nominal HR05 and HR09 resolution.
Micro-turbulence ($\xi$t) is derived using the relation in Tsantaki et al.~(2013) and macroturbulence is from Doyle et al.~(2014). We adopt a metallicity of [Fe/H] = 0.0. All atmospheric parameters were kept fixed in our analysis and we allowed only V$\sin$i to vary in our $\chi ^2$ minimisation procedure.

The synthetic spectra were computed with the plane-parallel, local thermodynamic equilibrium (LTE) code MOOG (Sneden~1973; Sneden et al.~2012). The model atmospheres were calculated with the {\tt ATLAS9}; starting from the grid of models available at F. Castelli's website\footnote{\url{http://www.oact.inaf.it/castelli/}}. When HR05 and HR09 spectra were available, we consider the final rotational velocity as the mean value measured from the two settings.

We calibrated the uncertainties of our $V\sin i$ measurements by using the stars with measurements in both gratings. When normalized by the squared sum of the uncertainties, the differences between the gratings should be normally distributed with a standard deviation of 1. To achieve this, a correction factor of $0.36$ had to be applied to all uncertainties.

In Fig.~\ref{fig:cmd_rotation} we show the CMD of cluster members along with the measured Vsini for the subsample of 60 members.  Light blue represents rapid rotators while dark blue/black shows more slow rotators.  It is evident that the rapid rotators preferentially lie to the red of the MSTO, whereas the slower rotators lie closer to the blue nominal main sequence.  This is consistent with observations of young (Dupree et al.~2017) and intermediate age (Kamann et al.~2018) massive clusters in the LMC/SMC.  We will quantitatively compare our observations with model predictions including rotation in \S~\ref{sec:comparison}.  The properties of cluster members with Vsini measurements are listed in Table~\ref{tab:objects}.

\subsection{Extinction towards the cluster}

The extinction towards the cluster was estimated by comparing the MIST (Choi et al.~2016) isochrones, at solar metallicity, to the observed CMD, at a fixed adopted distance (from the Gaia DR2 parallax).  We allowed age to be a free parameter (between 100~Myr and 3 Gyr), and concentrated the fit on matching the main sequence (i.e., a region largely independent of the adopted age).  We adopted the mean extinction coefficients for the Gaia filters from Casagrande \& Vandenberg~(2018), assuming $R_{\rm V}$ = 3.1.  $A_V$ was adjusted until the main sequence of the isochrones passed through the centre of the observed main sequence (between $16 \le G_{\rm BP} \le 18$).  We found that $A_V = 0.9 \pm0.1$~mag provided the best fit to the data.

The result, however, is metallicity dependent.  Performing the same test with models of [Fe/H]$=-0.5$ led to a value of $A_V = 1.3$~mag.  However, at this metallicity, the model isochrones displayed a curvature below the MSTO data that was not found in the observations.  Hence, we adopt the value determined from the solar metallicity isochrones.

\subsection{Cluster mass}

We estimate the mass of the cluster by counting the number of stars within 2 magnitude intervals that should not be heavily affected by incompleteness or stellar evolution.  We convert these magnitude intervals to mass, assuming an age of $800$~Myr (see below), distance and extinction described above, and compare the number of observed stars with expectations from stochastically sampling from a Kroupa~(2002) initial mass function.  We find 57 and 28 stars within the mass intervals of $1.39-1.68~\msun$ (G$_{\rm BP} = 16-17$~mag) and $1.68-1.85~\msun$  (G$_{\rm BP} = 15.5-16$~mag), respectively.  Comparing these numbers to $50,000$ realisations of clusters with masses between $500-5000$~\msun, we find that the cluster initially contained $2400\pm300$~\msun\ (not taking into account mass loss from tidal effects or two body relaxation).  

\section{Results}

\subsection{Extent of the eMSTO}

\begin{figure}
\centering
\includegraphics[width=8cm]{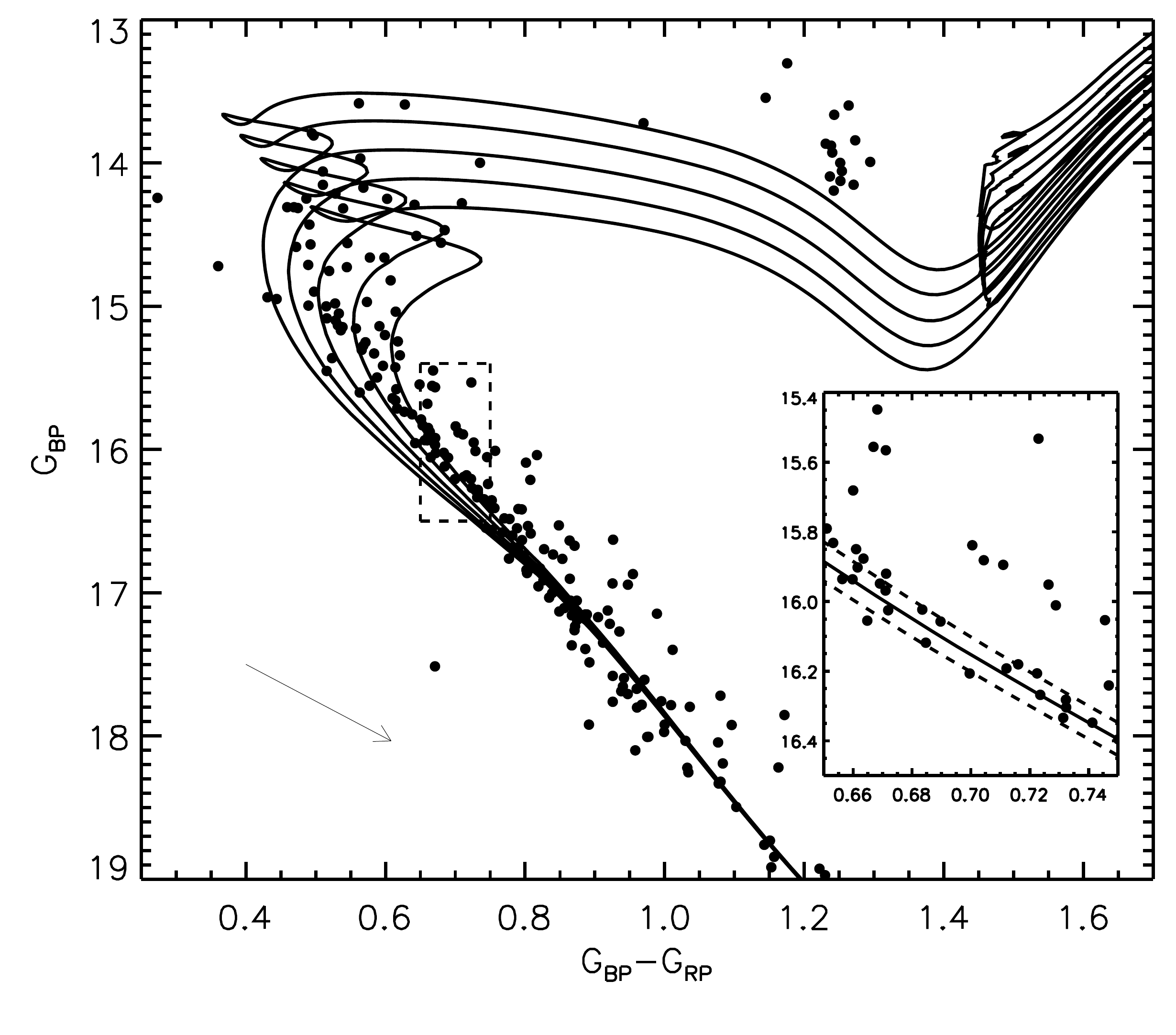}
\includegraphics[width=8cm]{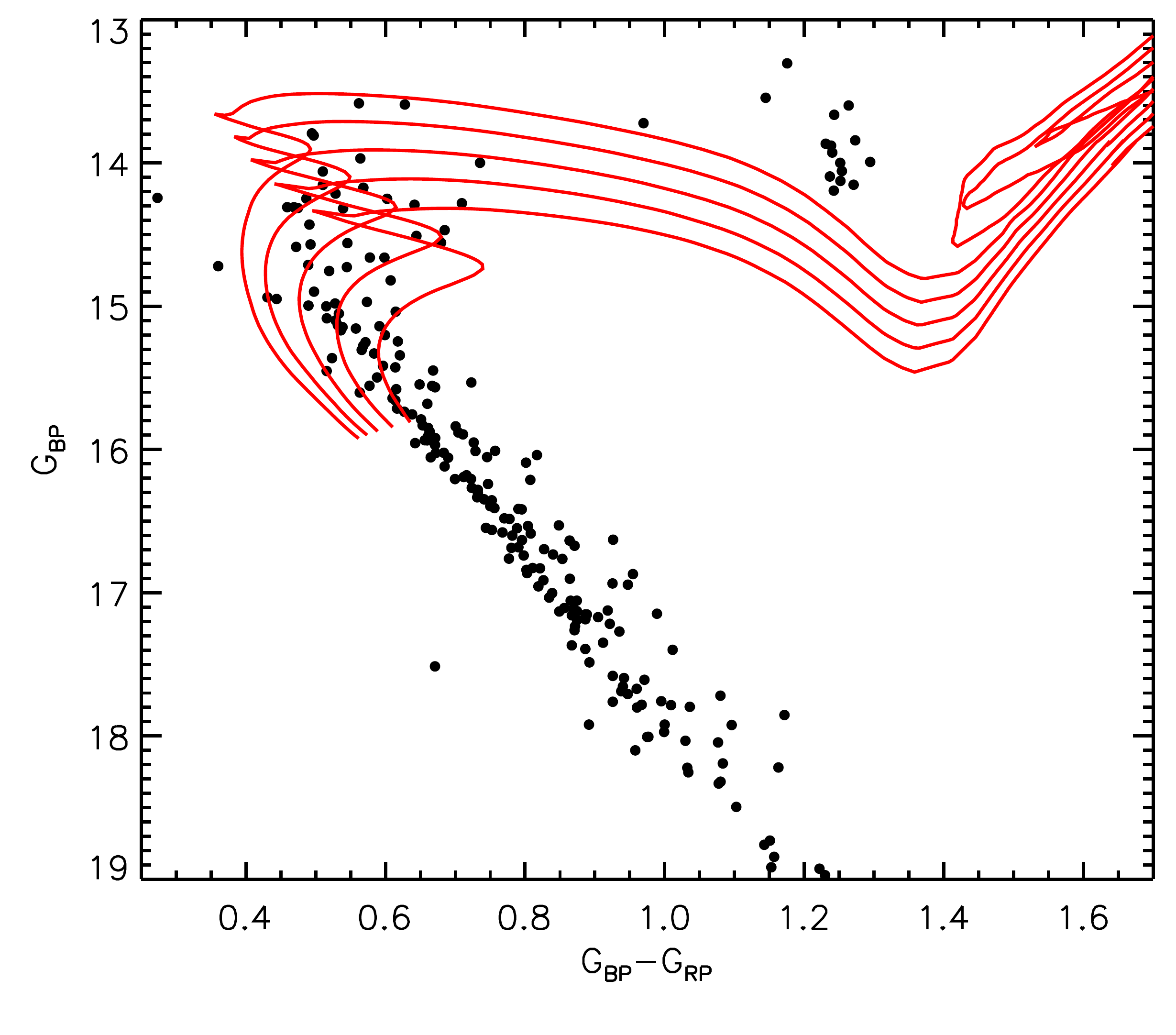}
\caption{{\bf Top panel:} The CMD (in Gaia passbands) of the cluster members based on the adopted selection criteria. The MIST isochrones (for standard models) for ages of log t = $8.75...8.95$ (in steps of 0.05) are overplotted. The arrow shows the effect of $A_V = 0.5$~mag of extinction in the CMD.  The dashed box is shown in more detail in the inset, where here the solid line is the nominal ridge line for the single star MS and the dashed lines show the same line shifted by 0.01 magnitudes in colour.  The narrowness of the MS indicates very little differential extinction is present.  A binary sequence above the MS is clearly visible.  {\bf Bottom panel:}  The same as the top panel but now for the Geneva isochrones with ages between log t = $8.85...9.05$ (in steps of 0.05), for an initial rotation rate of $\Omega / \Omega_{\rm c, initial} = 0.4$.  The lowest mass stars in the isochrone is 1.7$\msun$.}
\label{fig:cmd}
\end{figure}

As shown by Platais et al.~(2012) differential extinction can mimic an eMSTO if not properly taken into account.  A way to estimate the role of differential extinction is to look at the width of the main sequence.  An example region is shown in the inset of the upper panel of Fig.~\ref{fig:cmd}, where we have fit a nominal ridge line to this section of the main sequence (solid line).  We also show the same ridge line shifted by 0.01 magnitudes in colour in both directions, which encompasses most of the stars on the single star main sequence.   By applying different levels of extinction to the nominal ridge line (within the inset box in Fig.~\ref{fig:cmd}), we found that the amount of differential extinction present to be $A_V \leq 0.07$~mag, in order to stay within the 0.01~magnitude tolerance in colour.\footnote{This is smaller than the uncertainty in the absolute extinction towards the clusters because it is independent of the shape of the isochrone.} Additionally, we see a sequence of binary stars well separated from the main sequence.

In the upper panel of Fig.~\ref{fig:cmd} we overplot the MIST (Choi et al.~2016) isochrones for solar metallicity, and extinction value of $A_V = 0.9$, at the adopted distance.  We plot a range of ages from log t=8.75...8.95 (in steps of 0.05) that encompasses the width of the MSTO.  Extrapolating between the outer isochrones we see that an age of log t = 8.85 fits the main part of the eMSTO (in agreement with previous age estimates - Mermilliod et al.~2001) with an inferred age spread between log t=8.775 and 8.925 ($\sim245$~Myr).  More quantitatively, following on previous work (e.g., Goudfrooij et al.~2014;  Niederhofer et al. 2016) we estimated the age of each star on the MSTO ($0.4 \le G_{\rm BP} - G_{\rm RP} \le 0.8$, $14 \le G_{\rm BP} \le 15.5$) by finding the isochrone that passes closest to the star in the CMD (using isochrones with the same parameters as above and in steps of $\Delta(log age) = 0.01$).  The mean age found with this method is $775$~Myr with a dispersion of $120$~Myr.  We convert the estimated $\sigma$ to FWHM in order to more directly compare with previous estimates, finding FWHM$=280$~Myr, in good agreement with that found by eye.

In the bottom panel of Fig.~\ref{fig:cmd} we show the same analysis using the Geneva/SYCLIST isochrones using stellar models including rotation with an initial  rotation rate of $\Omega / \Omega_{\rm c, initial} = 0.4$ (Georgy et al.~2013a,b) with the same adopted parameters.  We find that the two sets of isochrones give similar results, with the Geneva models giving slightly older ages, with a best fit age of log t = 8.95 and a spread of $\sim300$~Myr.

\begin{figure}
\centering
\includegraphics[width=8.5cm]{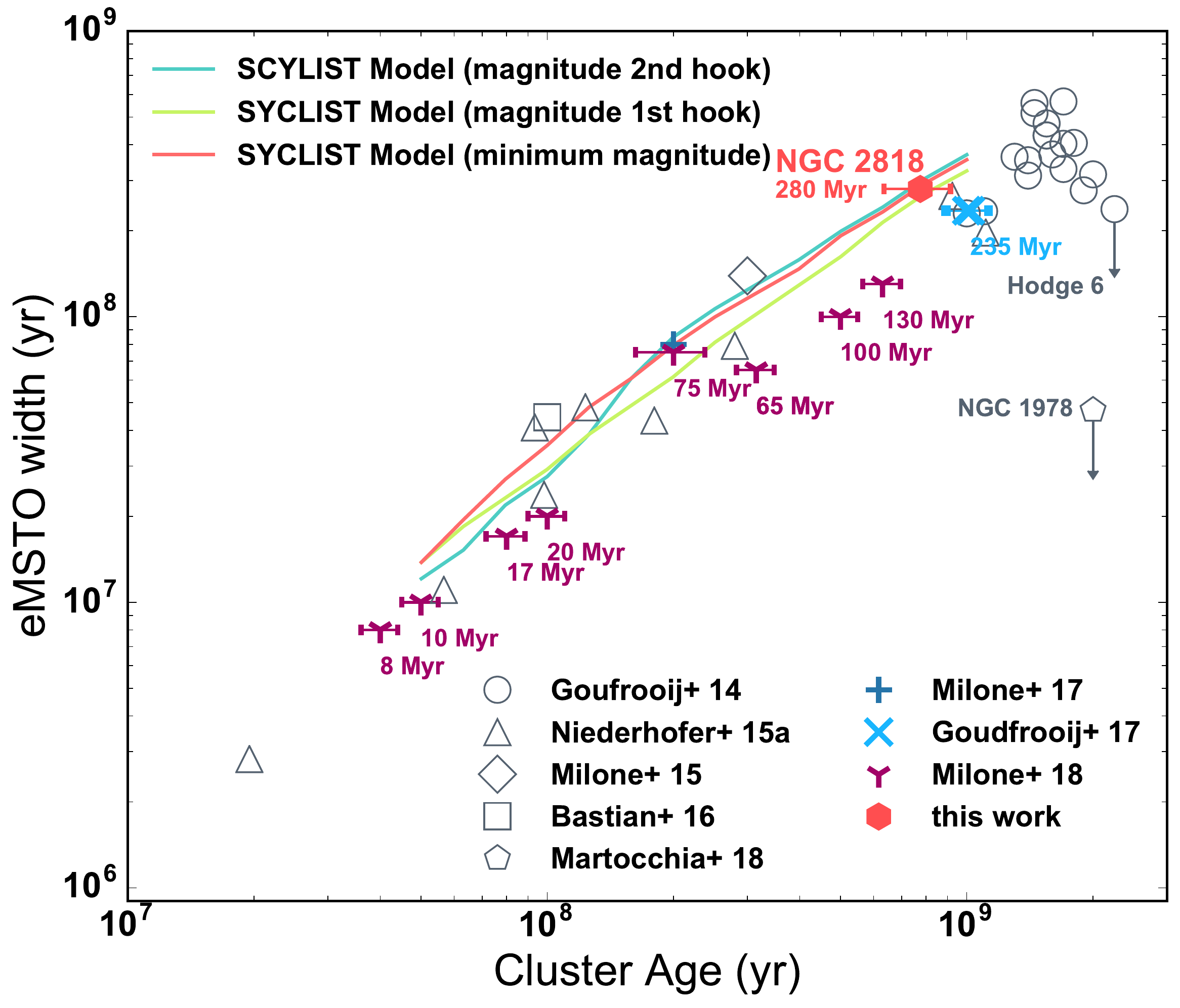}
\caption{The relation between cluster age and inferred age spread for a sample of clusters from the literature as well as that found for NGC~2818 (see Niederhofer et al.~2015 for the original plot and a description of the models).  We have added additional clusters from the literature in the past 3 years (see text).  NGC~2818 follows the trend found for other cluster samples.  The horizontal 'error bars' for the Goudfrooij et al.~(2017) and Milone et al.~(2018) samples show, assuming the inferred age spreads are real, when star formation began and terminated within each cluster.  Solid lines show the expectations for rotating stellar models (Niederhofer et al.~2015). }
\label{fig:age_delta}
\end{figure}

We can compare the estimated age and inferred age spread with previous results in the literature.  Niederhofer et al.~(2015) showed that there exists a tight relation, as derived for LMC/SMC clusters, between the age of the cluster and the inferred age spread on the MSTO.  The authors also showed that this is expected from models that included stellar rotation.  In Fig.~\ref{fig:age_delta} we show an updated version of the Niederhofer et al. figure and highlight the position of NGC~2818.  We also include recent results from Milone et al.~(2015, 2016, 2018), Bastian et al.~(2016), Goudfrooij et al.~(2017), and Martocchia et al.~(2018).  For the Milone et al.~(2018) results, we adopt the difference between the ages of the non-rotating isochrone that fits the blue sequence and the rotating isochrone ($\Omega=0.9$) that fits the red sequence.\footnote{Note that this is different from the method used in all other referenced works in the current paper, and likely results in smaller age ranges that the other methods (as can be seen in Fig.~10 of Milone et al.~2018).  This may explain why the Milone et al.~(2018) data points are systematically below that found for other studies.  However, a strong trend still exists between the age and inferred age spread for the clusters in their sample.}  For the Goudfrooij et al.~(2017) results we adopt the age difference between the two peaks of the pseudo-age distribution (their Fig.~9).  

The horizontal bars on the Milone et al.~(2018) and Goudfrooij et al.~(2017) represent the duration between when star-formation began and terminated, if the eMSTO is interpreted as an age spread. Note that if interpreted as an age spread, each cluster would begin and end to form stars at a different time.  It would then need to be a coincidence that we observed each cluster at a very specific age in order to get the trend between age and age-spread.

Lines show the expected trend from stellar rotation as predicted by the Geneva/SYCLIST models (details below).  The results for NGC~2818 nicely follow the same trend as previous results from the literature, although we note that the theoretical curves were calculated for LMC metallicity.

\subsection{Comparison with stellar models with rotation}
\label{sec:comparison}

\begin{figure}
\centering
\includegraphics[width=8cm]{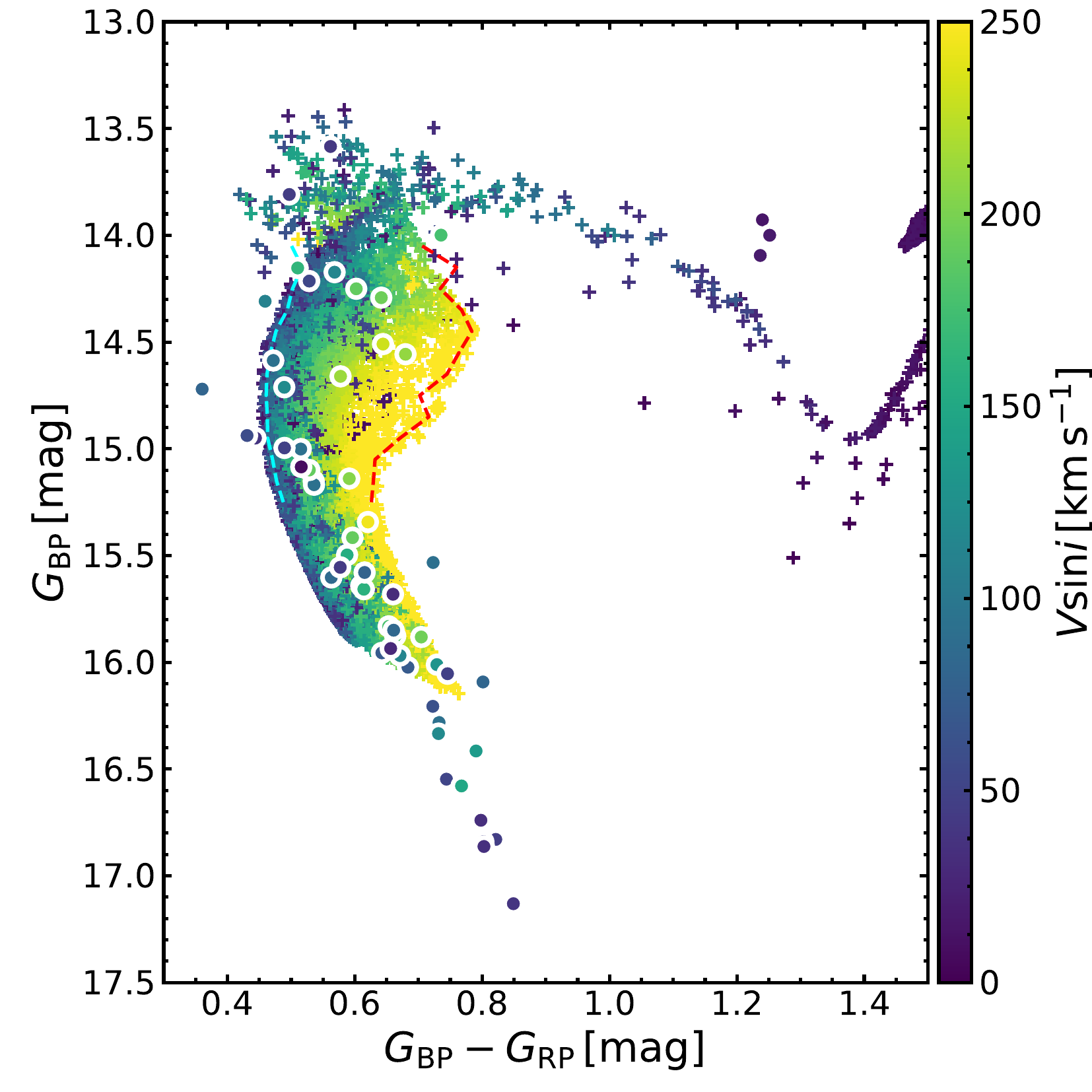}
\includegraphics[width=8cm]{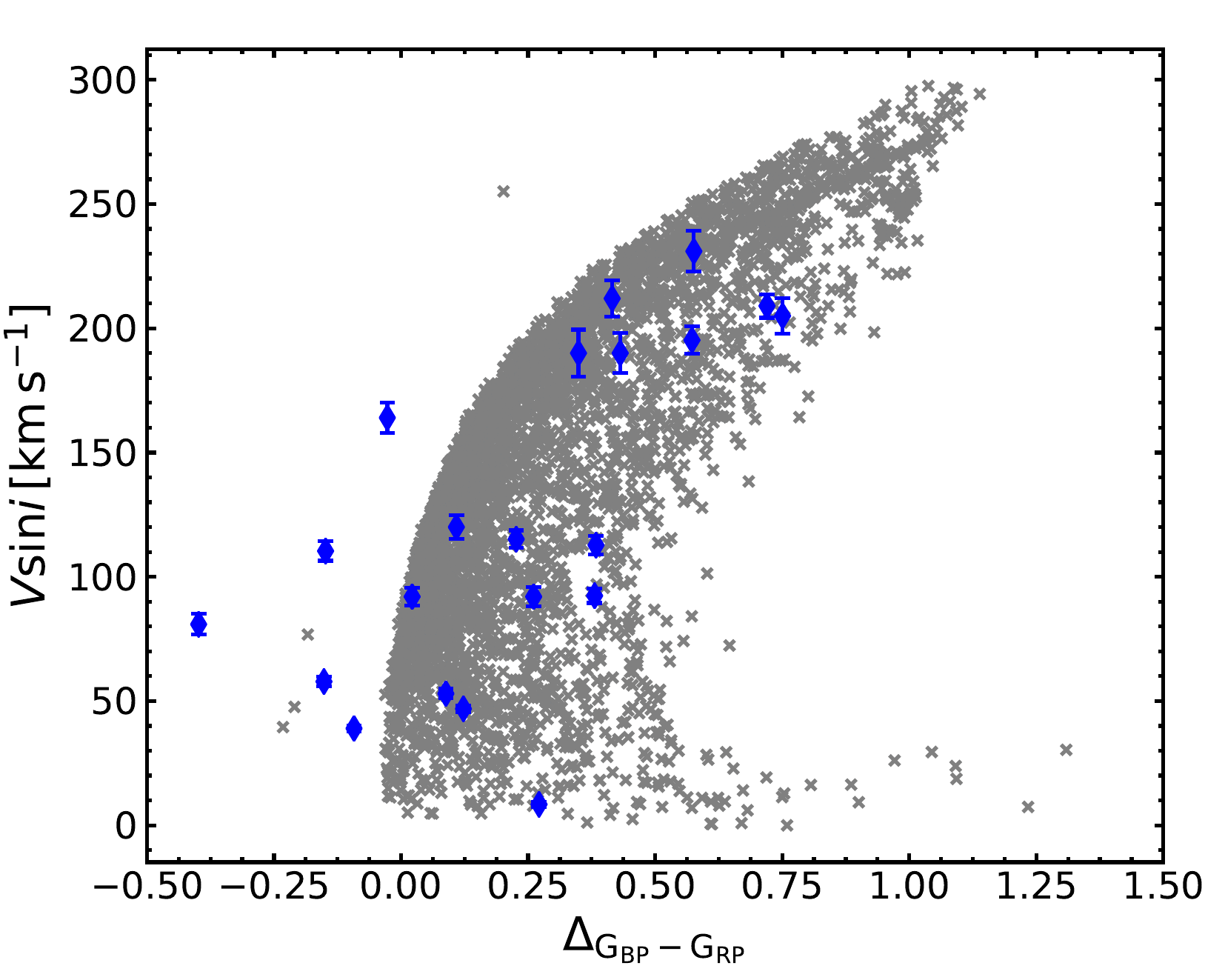}
\caption{{\bf Top panel:} The observed CMD of NGC~2818 members colour-coded by their measured Vsini values.  The background coloured crosses show the predictions from the SYCLIST models, also colour-coded by their Vsini values.  The adopted blue and red ridgelines are shown as dashed lines.  {\bf Bottom panel:} The Vsini value vs. the pseudo-colour (normalised colour difference from the blue ridgeline (top panel).  We have only included stars with magnitudes brighter than G$_{\rm BP} = 15.3$ (i.e. the MSTO) in the figure. }
\label{fig:comparison}
\end{figure}

In order to address whether stellar rotation is the underlying cause of the eMSTO in NGC~2818, we have compared the observed CMD and Vsini measurements to predictions from the SYCLIST synthetic cluster models.  For this we have assumed an age of log t (yr)$=8.95$, solar metallicity, a random inclination angle distribution, and a range of rotation rates from non-rotating to near break up ($\Omega/\Omega_{\rm c,initial}=0.95$).  The models include gravity darkening (Espinosa Lara \& Rieutord~2011) as well as geometric deformation at high rotational velocities (Georgy et al.~2014).  We then determined the 5th and 95th percentiles in colour in 0.1 magnitude bins in G$_{\rm BP}$ over the range $13.5 \leq$ G$_{\rm BP}$ $\leq 15.3$, to define blue and red ridge lines, respectively (see the top panel of Fig.~\ref{fig:comparison}).  We then measured the normalised distance (in colour) to the blue ridgeline for every star brighter than G$_{\rm BP} = 15.3$ (i.e. the MSTO).

The bottom panel of Fig.~\ref{fig:comparison} presents the results where the points with error bars correspond to the observed stars in NGC~2818, while the grey crosses show the prediction from the models.   At blue pseudo-colours, the observations show that the average Vsini is quite low and it significantly increases towards redder colours. This matches the expectations of the models.  

\section{Discussion and Conclusions}

\subsection{Mismatch between the observed and expected position of the red clump}

A notable feature in Fig.~\ref{fig:cmd} is the fact that the models do not reproduce the position of the red clump.  We found that this holds for the MIST, Geneva and Padova (Marigo et al.~2017) isochrones, independent of the metallicity adopted (even when allowing extinction to be a free parameter).  If the distance is left as a free parameter, in addition to the age and extinction, an acceptable fit could be found.  This suggests that there is either a problem in the calibration of the models for the red clump, or that the conversion between theoretical properties of the isochrones (temperature, gravity and luminosity) to observational space in Gaia filters is off. Further observations of more open clusters, at different ages, should be able to differentiate between these possibilities.

\subsection{The Origin of the Extended Main Sequence Phenomenon}

The discovery of an eMSTO in a low mass ($\sim2400$~\msun), intermediate age ($800-900$~Myr) Galactic open cluster shows that the eMSTO phenomenon is not limited to high mass clusters, nor the special conditions of the LMC/SMC.  Combining the clean CMD made possible from Gaia proper motions and parallaxes, with medium resolution spectroscopy, shows that rapidly rotating stars are preferentially found on the red (cool) side of the MSTO, while slow rotators are found on the blue, nominal MSTO.  Such a relation between colour and rotational velocity (Vsini) is expected from models that include stellar rotation. 

In a future work we will study a sample of open clusters available in the Gaia DR release to investigate the age - inferred age relation at solar metallicity (de Juan Ovelar et al. in prep.), and directly compare the observations to models that include magnetic breaking due to the onset of convection in lower mass stars (Georgy et al. in prep.).


\begin{table*} 

  \begin{tabular}
    {lllccccccccc} Gaia DR2 ID& RA & Dec & G$_{\rm BP}$  & G$_{\rm RP}$  & Vsini$_{\rm HR09}$ & eVsini$_{\rm HR09}$ & Vsini$_{\rm HR05}$ & eVsini$_{\rm HR05}$  & Vsini$_{\rm adopted}$ & eVsini$_{\rm adopted}$ \\
&  [deg] & [deg] & [mag] & [mag] & [mag] & [mag] & [km/s]  & [km/s] & [km/s]  & [km/s] \\
    \hline 
   5623333339867028096  &  139.044769287 &   -36.650726318  & 15.68 & 15.02 &  31.0 &   2.8 &   0.0 &   0.0 &  31.0 &   2.8\\
 5623333305507502720  &  139.018203735 &   -36.63874054   & 13.81 & 13.31 &  44.0 &   4.9 &  44.0 &   4.0 &  44.0 &   3.6\\
 5623339318459536384  &  139.059326172 &   -36.609886169  & 14.95 & 14.51 &   0.0 &   0.0 &  39.0 &   3.7 &  39.0 &   3.7\\
 5623380378350117504  &  139.036895752 &   -36.613384247  & 13.93 & 12.69 &  14.0 &   1.5 &   0.0 &   0.0 &  14.0 &   1.5\\
 5623380442769067520  &  139.01902771  &    -36.597003937 & 14.09 & 12.86 &  19.0 &   1.2 &  17.0 &   0.9 &  17.7 &   0.9\\
 5623386528743281408  &  139.012268066 &   -36.576229095  & 14.00 & 12.75 &  17.0 &   1.2 &   0.0 &   0.0 &  17.0 &   1.2\\
 5623339352821216896  &  139.063354492 &   -36.600475311  & 15.53 & 14.81 &  91.0 &   7.3 &   0.0 &   0.0 &  91.0 &   7.3\\
 5623333339867029120  &  139.040908813 &   -36.650588989  & 14.56 & 13.88 & 224.0 &  22.6 & 201.0 &  16.5 & 209.0 &  15.5\\
 5623332858830717440  &  138.984985352 &   -36.673389435  & 14.51 & 13.87 & 231.0 &  26.9 &   0.0 &   0.0 & 231.0 &  26.9\\
 5623386631822488832  &  139.025756836 &   -36.554733276  & 16.28 & 15.55 &  92.0 &   8.0 &   0.0 &   0.0 &  92.0 &   8.0\\
 5623386563103010560  &  139.041259766 &   -36.562767029  & 14.00 & 13.26 & 187.0 &  19.0 & 170.0 &  14.1 & 176.0 &  13.2\\
 5623380786366456448  &  138.928314209 &   -36.62758255   & 14.29 & 13.65 & 204.0 &  25.4 & 190.0 &  19.6 & 195.2 &  18.1\\
 5623339249742007808  &  139.060165405 &   -36.615459442  & 16.83 & 16.02 &  42.0 &   2.8 &  40.0 &   2.8 &  41.0 &   2.3\\
 5623338974864103808  &  139.083496094 &   -36.642276764  & 16.83 & 16.01 &  46.0 &   3.7 &  44.0 &   4.0 &  45.1 &   3.2\\
 5623386391304315904  &  139.065444946 &   -36.570781708  & 15.15 & 14.62 & 118.0 &  14.4 & 107.0 &  15.0 & 112.7 &  12.3\\
 5623339455900419840  &  139.080383301 &   -36.585018158  & 14.21 & 13.69 &  53.0 &   6.1 &   0.0 &   0.0 &  53.0 &   6.1\\
 5623339455900421248  &  139.080383301 &   -36.588939667  & 14.25 & 13.65 & 190.0 &  26.3 &   0.0 &   0.0 & 190.0 &  26.3\\
 5623339176724479232  &  139.094100952 &   -36.604549408  & 16.74 & 15.94 &  34.0 &   3.4 &   0.0 &   0.0 &  34.0 &   3.4\\
 5623386253865372288  &  139.045028687 &   -36.598823547  & 14.59 & 14.11 &   0.0 &   0.0 &  92.0 &  11.6 &  92.0 &  11.6\\
 5623333442946228992  &  139.043182373 &   -36.61378479   & 13.58 & 13.02 &   0.0 &   0.0 &  41.0 &   4.0 &  41.0 &   4.0\\
 5623380270970381696  &  138.994445801 &   -36.629962921  & 16.55 & 15.80 &  51.0 &   3.7 &   0.0 &   0.0 &  51.0 &   3.7\\
 5623333408586502656  &  139.035614014 &   -36.636032104  & 16.84 & 16.04 &  49.0 &   4.0 &  43.0 &   3.7 &  45.8 &   3.2\\
 5623339077943312128  &  139.082473755 &   -36.624046326  & 15.88 & 15.21 & 202.0 &  22.3 & 173.0 &  18.0 & 184.5 &  16.5\\
 5623339284101742592  &  139.076309204 &   -36.619358063  & 15.42 & 14.82 &   0.0 &   0.0 & 191.0 &  22.3 & 191.0 &  22.3\\
 5623332309074877312  &  139.074935913 &   -36.679534912  & 15.00 & 14.49 &  92.0 &  12.8 &   0.0 &   0.0 &  92.0 &  12.8\\
 5623332205995667456  &  139.064376831 &   -36.684459686  & 15.17 & 14.63 &  96.0 &  12.8 &  90.0 &  10.4 &  92.4 &   9.5\\
 5623332996269657984  &  139.033828735 &   -36.681102753  & 15.83 & 15.18 &   0.0 &   0.0 & 177.0 &  16.2 & 177.0 &  16.2\\
 5623380928105950720  &  138.946029663 &   -36.6171875    & 15.00 & 14.51 &  48.0 &   5.5 &  46.0 &   4.9 &  46.9 &   4.3\\
 5623381340422800512  &  138.958129883 &   -36.58423996   & 14.94 & 14.51 &  59.0 &   8.6 &  57.0 &   7.3 &  57.8 &   6.6\\
 5623386597462756480  &  139.006134033 &   -36.56627655   & 15.50 & 14.91 & 146.0 &  16.8 & 171.0 &  22.3 & 155.1 &  15.7\\
 5623386597462754176  &  139.005584717 &   -36.557128906  & 15.64 & 15.03 & 159.0 &  18.4 & 144.0 &  19.9 & 152.1 &  16.0\\
 5623386494383526400  &  139.068847656 &   -36.548202515  & 15.60 & 15.04 &  89.0 &   9.8 &  83.0 &   8.6 &  85.6 &   7.6\\
 5623386837980904832  &  139.094543457 &   -36.540142059  & 16.01 & 15.28 & 133.0 &  11.0 & 124.0 &  12.5 & 129.1 &   9.8\\
 5623338081510889344  &  139.144302368 &   -36.651504517  & 16.02 & 15.34 &   0.0 &   0.0 &  71.0 &   5.8 &  71.0 &   5.8\\
 5623338253309593728  &  139.103302002 &   -36.649654388  & 15.14 & 14.55 &   0.0 &   0.0 & 205.0 &  23.6 & 205.0 &  23.6\\
 5623380756307269120  &  138.918502808 &   -36.643554688  & 16.58 & 15.81 & 149.0 &   9.8 &   0.0 &   0.0 & 149.0 &   9.8\\
 5623387422096548864  &  138.98815918  &    -36.52878952  & 15.66 & 15.04 & 171.0 &  21.1 & 152.0 &  17.4 & 159.7 &  15.8\\
 5623387898833928448  &  139.043746948 &   -36.506778717  & 14.66 & 14.08 &   0.0 &   0.0 & 212.0 &  23.9 & 212.0 &  23.9\\
 5623387078499074688  &  139.065994263 &   -36.520988464  & 15.88 & 15.18 & 196.0 &  18.7 &   0.0 &   0.0 & 196.0 &  18.7\\
 5623386941060117760  &  139.107223511 &   -36.531391144  & 15.58 & 14.96 &  83.0 &   8.3 &  76.0 &   7.0 &  78.9 &   6.3\\
 5623386941060118656  &  139.104309082 &   -36.534217834  & 16.21 & 15.48 &  73.0 &   5.5 &  55.0 &   4.6 &  62.4 &   4.2\\
 5623338837425108864  &  139.16394043  &    -36.59770584  & 15.85 & 15.19 &  92.0 &   8.0 &  78.0 &   6.4 &  83.5 &   5.9\\
 5623338528187471616  &  139.170883179 &   -36.626014709  & 16.42 & 15.62 &   0.0 &   0.0 & 136.0 &   9.5 & 136.0 &   9.5\\
 5623380515789076992  &  138.987854004 &   -36.599262238  & 16.86 & 16.06 &  44.0 &   2.8 &  30.0 &   1.8 &  34.3 &   1.8\\
 5623379824294763008  &  138.995162964 &   -36.658473969  & 17.13 & 16.28 &  40.0 &   3.7 &  37.0 &   3.1 &  38.2 &   2.8\\
 5623380481429346304 &  138.972946167 &   -36.616920471   & 14.17 & 13.60 & 120.0 &  15.6 & 112.0 &  12.5 & 115.1 &  11.5\\
 5623332893190448384  &  138.997970581 &   -36.663852692  & 14.31 & 13.85 & 116.0 &  17.7 & 107.0 &  13.8 & 110.4 &  12.7\\
 5623380550148822144  &  138.969726562 &   -36.610404968  & 14.15 & 13.64 &   0.0 &   0.0 & 164.0 &  20.2 & 164.0 &  20.2\\
 5623380378350116864  &  139.027633667 &   -36.602481842  & 15.96 & 15.31 &  71.0 &   6.1 &   0.0 &   0.0 &  71.0 &   6.1\\
 5623380172191705600  &  138.955978394 &   -36.619419098  & 14.71 & 14.22 &   0.0 &   0.0 & 120.0 &  15.6 & 120.0 &  15.6\\
 5623380309630647552  &  139.004837036 &   -36.616539001  & 16.09 & 15.29 &   0.0 &   0.0 &  82.0 &   5.2 &  82.0 &   5.2\\
 5623380275270915840  &  138.986404419 &   -36.628726959  & 15.56 & 14.98 &   0.0 &   0.0 &  43.0 &   4.3 &  43.0 &   4.3\\
 5623380412709857920  &  139.018722534 &   -36.610397339  & 15.10 & 14.57 & 190.0 &  30.9 &   0.0 &   0.0 & 190.0 &  30.9\\
 5623380378350115840  &  139.033843994 &   -36.605182648  & 15.08 & 14.57 &   7.0 &   4.9 &   9.0 &   3.1 &   8.4 &   3.0\\
 5623333305507291392  &  139.027664185 &   -36.642311096  & 14.72 & 14.36 &  80.9 &  13.5 &   0.0 &   0.0 &  80.9 &  13.5\\
 5623380378350114816  &  139.034606934 &   -36.601577759  & 16.05 & 15.31 &  48.0 &   4.0 &  46.0 &   3.7 &  46.9 &   3.2\\
 5623333236787821312  &  139.031326294 &   -36.661888123  & 15.34 & 14.72 & 245.0 &  31.2 &   0.0 &   0.0 & 245.0 &  31.2\\
 5623379828594324608  &  138.987945557 &   -36.666103363  & 16.33 & 15.60 & 127.0 &  11.0 & 113.0 &   9.5 & 119.0 &   8.5\\
 5623333133708811520  &  139.047271729 &   -36.655601501  & 15.97 & 15.30 & 114.0 &   9.8 &   0.0 &   0.0 & 114.0 &   9.8\\
 5623333271147560960  &  139.010131836 &   -36.647571564  & 15.94 & 15.28 &  38.0 &   3.1 &  26.0 &   1.8 &  29.2 &   1.8\\

    \hline 
  \end{tabular}
\caption{Properties of the stars with Vsini measurements.  The magnitudes are for the Gaia blue band pass (G$_{\rm BP}$) and Gaia red band pass (G$_{\rm RP}$).  We list three Vsini values, Vsini$_{\rm HR09}$ is that determined from the HR09 grating, Vsini$_{\rm HR05}$ is from the HR05 grating, and the V$_{\rm adopted}$ is the adopted value.  '0.0' values denote stars where a measurement was possible in that grating. Errors on the photometry are not given, as they are generally $<0.01$~mag.}
\label{tab:objects}
\end{table*}

\section*{Acknowledgments}

We thank Carmela Lardo for her help in the Vsini measurements and Elena Pancino and Peter Stetson for bringing NGC~2818 to our attention.  The anonymous referee is thanked for helpful suggestions that improved the manuscript.   NB gratefully acknowledges financial support from the Royal Society (University Research Fellowship).  NB, SK, and CU gratefully acknowledges financial support from the European Research Council (ERC-CoG-646928, Multi-Pop).  CC acknowledges financial support from the Swiss National Science Foundation (SNF). CC and CG thank the Equal Opportunity Office of the University of Geneva.  Support for this work was provided by NASA through Hubble Fellowship grant HST-HF2-51387.001-A awarded by the Space Telescope Science Institute, which is operated by the Association of Universities for Research in Astronomy, Inc., for NASA, under contract NAS5-26555.  This work has made use of data from the European Space Agency (ESA) mission {\it Gaia} (\url{https://www.cosmos.esa.int/gaia}), processed by the {\it Gaia} Data Processing and Analysis Consortium (DPAC, \url{https://www.cosmos.esa.int/web/gaia/dpac/consortium}). Funding for the DPAC has been provided by national institutions, in particular the institutions participating in the {\it Gaia} Multilateral Agreement.

\bsp
\label{lastpage}

\vspace{-0.5cm}
\begin{thebibliography}{99}

\bibitem[Bastian \& de Mink(2009)]{BastianDeMink09}
Bastian, N., \& de Mink, S. E. 2009,
MNRAS, 398, L11

\bibitem[Bastian et al.(2016)]{2016MNRAS.460L..20B} Bastian, N., Niederhofer, F., Kozhurina-Platais, V., et al.\ 2016, MNRAS, 460, L20 

\bibitem[Bastian \& Lardo(2017)]{2017arXiv171201286B} Bastian, N., \& Lardo, C.\ 2018, ARA\&A, v56, in press (arXiv:1712.01286)

\bibitem[Bessell et al.(1998)]{bessell98} Bessell, M.~S., Castelli, F., \& Plez, B.\ 1998, A\&A, 333, 231 

\bibitem[Brandt \& Huang(2015)]{2015ApJ...807...24B} Brandt, T.~D., \& Huang, C.~X.\ 2015a, ApJ, 807, 24 

\bibitem[Brandt \& Huang(2015)]{2015ApJ...807...25B} Brandt, T.~D., \& Huang, C.~X.\ 2015b, ApJ, 807, 25 

\bibitem[Cabrera-Ziri et al.(2016)]{2016MNRAS.457..809C} Cabrera-Ziri, I., Bastian, N., Hilker, M., et al.\ 2016, MNRAS, 457, 809 

\bibitem[Casagrande \& VandenBerg(2018)]{2018MNRAS.479L.102C} Casagrande, L., \& VandenBerg, D.~A.\ 2018, MNRAS, 479, L102 

\bibitem[Choi et al.(2016)]{2016ApJ...823..102C} Choi, J., Dotter, A., Conroy, C., et al.\ 2016, ApJ, 823, 102 

\bibitem[Conroy \& Spergel(2011)]{2011ApJ...726...36C} Conroy, C., \& Spergel, D.~N.\ 2011, ApJ, 726, 36 

\bibitem[Doyle et al.(2014)]{doyle14} Doyle, A.~P., Davies, G.~R., Smalley, B., Chaplin, W.~J., \& Elsworth, Y.\ 2014, MNRAS, 444, 3592

\bibitem[Dupree et al.(2017)]{dupree17} Dupree, A.~K., Dotter, A., Johnson, C.~I., et al.\ 2017, ApJL, 846, L1 

\bibitem[Espinosa Lara \& Rieutord(2011)]{2011A&A...533A..43E} Espinosa Lara, F., \& Rieutord, M.\ 2011, A\&A, 533, A43 

\bibitem[Flower(1996)]{flower96} Flower, P.~J.\ 1996, ApJ, 469, 355 

\bibitem[Gaia Collaboration et al.(2016)]{2016A&A...595A...1G} Gaia Collaboration, Prusti, T., de Bruijne, J.~H.~J., et al.\ 2016, A\&A, 595, A1 

\bibitem[Gaia Collaboration et al.(2018)]{2018arXiv180409365G} Gaia Collaboration, Brown, A.~G.~A., Vallenari, A., et al.\ 2018, A\&A, in press (arXiv:1804.09365) 


\bibitem[Georgy et al.(2013a)]{Georgy13} Georgy, C., Ekstr\"om, S., Granada, A. et al. 2013a, A\&A, 553, A24

\bibitem[Georgy et al.(2013b)]{Georgy13b} Georgy, C., Ekstr\"om, S., Eggenberger, P. et al. 2013b, A\&A, 558, 103

\bibitem[Georgy et al.(2014)]{2014A&A...566A..21G} Georgy, C., Granada, A., Ekstr{\"o}m, S., et al.\ 2014, A\&A, 566, A21 
 
\bibitem[Georgy et al.(2018)]{temp} Georgy, C., et al. A\&A, 2018, submitted

\bibitem[Goudfrooij et al.(2014)]{2014ApJ...797...35G} Goudfrooij, P., Girardi, L., Kozhurina-Platais, V., et al.\ 2014, ApJ, 797, 35 

\bibitem[Goudfrooij et al.(2017)]{2017ApJ...846...22G} Goudfrooij, P., Girardi, L., \& Correnti, M.\ 2017, ApJ, 846, 22 

\bibitem[Kamann et al.(2018)]{kamann18} Kamann, S. et al.~2018, MNRAS, submitted

\bibitem[Kroupa(2002)]{2002Sci...295...82K} Kroupa, P.\ 2002, Science, 295, 82 

\bibitem[Lindegren et al.(2018)]{2018arXiv180409366L} Lindegren, L., Hernandez, J., Bombrun, A., et al.\ 2018, A\&A in press (arXiv:1804.09366)

\bibitem[Mackey \& Broby Nielsen(2007)]{MackeyBrobyNielsen07}  Mackey, A.~D., \& Broby Nielsen, P.\ 2007, MNRAS, 379, 151 

\bibitem[Marigo et al.(2017)]{2017ApJ...835...77M} Marigo, P., Girardi, L., Bressan, A., et al.\ 2017, ApJ, 835, 77 

\bibitem[Marino et al.(2018)]{2018arXiv180704493M} Marino, A.~F., Przybilla, N., Milone, A.~P., et al.\ 2018, AJ, in press (arXiv:1807.04493)

\bibitem[Martocchia et al.(2018)]{2018MNRAS.477.4696M} Martocchia, S., Niederhofer, F., Dalessandro, E., et al.\ 2018, MNRAS, 477, 4696 

\bibitem[Mermilliod et al.(2001)]{2001A&A...375...30M} Mermilliod, J.-C., Clari{\'a}, J.~J., Andersen, J., Piatti, A.~E., \& Mayor, M.\ 2001, A\&A, 375, 30 

\bibitem[Milone et 
al.(2009)]{2009A&A...497..755M} Milone, A.~P., Bedin, L.~R., Piotto, G., \& Anderson, J.\ 2009, A\&A, 497, 755 

\bibitem[Milone et al.(2015)]{2015MNRAS.450.3750M} Milone, A.~P., Bedin, L.~R., Piotto, G., et al.\ 2015, MNRAS, 450, 3750 

\bibitem[Milone et al.(2016)]{2016MNRAS.458.4368M} Milone, A.~P., Marino, A.~F., D'Antona, F., et al.\ 2016, MNRAS, 458, 4368 

\bibitem[Milone et al.(2018)]{2018MNRAS.477.2640M} Milone, A.~P., Marino, A.~F., Di Criscienzo, M., et al.\ 2018, MNRAS, 477, 2640 

\bibitem[Niederhofer et al.(2015b)]{2015MNRAS.453.2070N} Niederhofer, F., 
Georgy, C., Bastian, N., \& Ekstr{\"o}m, S.\ 2015, MNRAS, 453, 2070 

\bibitem[Niederhofer et al.(2016)]{2016A&A...586A.148N} Niederhofer, F., Bastian, N., Kozhurina-Platais, V., et al.\ 2016, A\&A, 586, A148 

\bibitem[Piatti \& Bastian(2016)]{2016A&A...590A..50P} Piatti, A.~E., \& Bastian, N.\ 2016, A\&A, 590, A50 

\bibitem[Platais et al.(2012)]{2012ApJ...751L...8P} Platais, I., Melo, C., Quinn, S.~N., et al.\ 2012, ApJL, 751, L8 

\bibitem[Sneden(1973)]{sneden73} Sneden, C.\ 1973, ApJ, 184, 839 

\bibitem[Sneden et al.(2012)]{sneden12} Sneden, C., Bean, J., Ivans, I., Lucatello, S., \& Sobeck, J.\ 2012, Astrophysics Source Code Library, ascl:1202.009 

\bibitem[Stetson(2000)]{2000PASP..112..925S} Stetson, P.~B.\ 2000, PASP, 112, 925 

\bibitem[Torres(2010)]{torres10} Torres, G.\ 2010, AJ, 140, 1158 

\bibitem[Tsantaki et al.(2013)]{tsantaki13} Tsantaki, M., Sousa, S.~G., Adibekyan, V.~Z., et al.\ 2013, A\&A, 555, A150 

\bibitem[Twarog(1983)]{1983ApJ...267..207T} Twarog, B.~A.\ 1983, ApJ, 267, 207 

\bibitem[Twarog et al.(2015)]{2015AJ....150..134T} Twarog, B.~A., Anthony-Twarog, B.~J., Deliyannis, C.~P., \& Thomas, D.~T.\ 2015, AJ, 150, 134 





\end{thebibliography}
\end{document}